\documentclass{optica-article}

\journal{opticajournal} 

\articletype{Research Article}
\usepackage{soul}
\usepackage{lineno}

\begin{document}

\title{Ultrafast optical coherence elastography for volumetric and dynamic in vivo imaging}

\author{Yongkang Zhao,\authormark{1} and Guo-Yang Li\authormark{1,*}}

\address{\authormark{1}School of Mechanics and Engineering Science, Peking University, Beijing 100871, PR China}

\email{\authormark{*}lgy@pku.edu.cn} 


\begin{abstract*}
Imaging the mechanical properties of biological tissues \emph{in vivo} with high spatial and temporal resolution is essential for understanding physiological function and disease progression. Optical coherence elastography (OCE) provides label-free, micrometer-scale mapping of tissue biomechanics, but its application to dynamic and volumetric measurements has been limited by slow acquisition speeds and susceptibility to motion artifacts. Here we introduce ultrafast optical coherence elastography (ultrafast OCE), a general framework for real-time volumetric biomechanical imaging \emph{in vivo}. By combining synchronized multi-phase acquisition with a demodulation strategy intrinsically robust to spectral aliasing, ultrafast OCE decouples mechanical excitation from acquisition speed, enabling reconstruction of full wave fields from only three sequential B-mode images. The method achieves frame rates up to two orders of magnitude higher than conventional approaches while preserving high sensitivity over frequencies ranging from the acoustic to ultrasonic regimes. We further develop a motion-correction strategy that compensates for bulk tissue motion under physiological conditions. We validate ultrafast OCE in dynamically stretched phantoms and pulsatile arteries and demonstrate sub-second volumetric imaging of the cornea and skin \emph{in vivo}. Ultrafast OCE enables real-time interrogation of tissue biomechanics across space and time, opening new opportunities for mechanobiology, cardiovascular research, and clinical diagnostics.
\end{abstract*}

\section{Introduction}

Mechanical properties of soft biological tissues play a central role in regulating physiological function and disease progression, influencing processes such as cell mechanotransduction~\cite{RN1397,RN1612}, tissue remodeling~\cite{RN1514,RN1708}, and vascular dynamics~\cite{RN1498,RN1741}. The ability to image tissue biomechanics \emph{in vivo} with high spatial resolution and without exogenous labels is therefore of fundamental importance for both basic research and clinical translation~\cite{RN1708,RN1672,RN1739,RN1156,RN1536,RN81}. However, existing techniques remain limited by fundamental trade-offs among spatial resolution, imaging speed, and mechanical sensitivity~\cite{RN1364,RN1328,RN1693,RN99,RN1520}.

Optical coherence elastography (OCE), an extension of optical coherence tomography (OCT), probes tissue mechanics by tracking mechanically induced wave propagation~\cite{RN1441,RN39,RN464}. Owing to the micrometer-scale resolution of OCT, OCE provides substantially higher spatial resolution than ultrasound shear wave elastography~\cite{RN1498,RN453} and magnetic resonance elastography~\cite{RN1690,RN875}, enabling the visualization of microscale mechanical heterogeneity in biological tissues~\cite{RN884,RN469,RN98,RN99}.

Despite these advantages, OCE remains fundamentally limited by the speed of wave field acquisition, hindering dynamic biomechanical and volumetric imaging \emph{in vivo}. This limitation arises because most OCE implementations reconstruct wave fields using sequential scanning schemes, such as M--B scans~\cite{RN39,RN1441}. As a result, two-dimensional wave field imaging is typically limited to frame rates of approximately 1~Hz~\cite{RN337,RN469}, making the measurements highly susceptible to physiological motion and incapable of resolving rapid biomechanical processes. Consequently, most \emph{in vivo} OCE studies are confined to quasi-static conditions or simplified geometries, preventing real-time volumetric imaging of tissue biomechanics.

Recent advances in MHz swept sources, such as Fourier domain mode-locked (FDML) lasers, have substantially increased OCT acquisition speed and enabled high-speed B--M scanning strategies~\cite{RN1742,RN464,RN1756}. However, existing implementations generally reconstruct wave fields without synchronizing mechanical excitation and image acquisition. Consequently, the imaging frame rate must substantially exceed the excitation frequency, restricting excitation to the sub-kHz regime and ultimately limiting the spatial resolution of OCE~\cite{RN1231,RN1441}.

To overcome these limitations, we introduce ultrafast optical coherence elastography (ultrafast OCE), a general framework for real-time volumetric biomechanical imaging \emph{in vivo}. Ultrafast OCE combines synchronized multi-phase acquisition with an aliasing-robust demodulation strategy to decouple mechanical excitation from acquisition speed, enabling full wave field reconstruction from only three sequential B-scans. A motion-correction strategy further compensates for bulk tissue motion under physiological conditions. We validate ultrafast OCE using dynamically stretched phantoms, pulsatile arteries, and volumetric \emph{in vivo} imaging of the cornea and skin, demonstrating real-time biomechanical imaging in living systems.

\section{Materials and Methods}

\subsection{Swept-source optical coherence tomographic vibrography}

Consider a point scatterer located at $(0,0,z_0)$ with reflectance amplitude $r_{z_0}$ and axial displacement $\delta_{z_0}\sin(\omega_m t+\varphi_{z_0})$, where $\delta_{z_0}$ is the vibration amplitude, $\omega_m = 2\pi f_m$ is the angular frequency, $t$ denotes time, and $\varphi_{z_0}$ is the initial phase.

During the $j$-th ($j=1,2,3,\ldots$) optical coherence vibrography acquisition, starting at time $t_j-T/2$, a swept source with optical power $P(\hat{t})$ and linearly tuned wavenumber $k=k_0+k_1\hat{t}$ is used, where $-T/2\le\hat{t}<T/2$ and $1/T = f_a$ denotes the laser sweep rate. The OCT interferometric signal can therefore be written as ($k_1\hat{t}\ll k_0$)~\cite{RN39}
\begin{equation}
I_j(\hat{t})=
r_{z_0}P(\hat{t})
\cos\!\Big[
2k_0z_0+2k_1\hat{t}z_0+2k_0\delta_{z_0}\sin(\omega_m t+\varphi_{z_0})
\Big],
\label{eq:photocurrent}
\end{equation}
with $t=t_j+\hat{t}$.

Expanding Eq.~\eqref{eq:photocurrent} using the generalized Jacobi--Anger identity~\cite{Kuklinski2019PropertiesOG} yields
\begin{equation}
I_j(\hat{t})=
r_{z_0}P(\hat{t})
\sum_{n=-\infty}^{\infty}
J_n(2k_0\delta_{z_0})
\cos\!\left[
(2k_1z_0+n\omega_m)\hat{t}+2k_0z_0+n\varphi_{z_0}^j
\right],
\end{equation}
where $J_n(\cdot)$ denotes the Bessel function of the first kind and $\varphi_{z_0}^j=\varphi_{z_0}+\omega_m t_j$.

Assuming a Gaussian optical power envelope with a full width at half maximum (FWHM) of $\sigma T$, $P(\hat{t})=\exp\!\left[-4\ln2\,\frac{\hat{t}^2}{(\sigma T)^2}\right]$, the Fourier transform of the photocurrent signal can be written as
\begin{equation}\label{eq:photocurrent_FT}
F_j(\omega,\omega_0)=
\frac{r_{z_0}}{2}
\left(\sqrt{\frac{\pi}{\ln2}}\frac{\sigma T}{2}\right)
\sum_{n=-\infty}^{\infty}
J_n(2k_0\delta_{z_0})
e^{i(2k_0z_0+n\varphi_{z_0}^j)}
\exp\!\left[
-\ln2\,\frac{(\omega-\omega_0-n\omega_m)^2}{(\Delta\omega)^2}
\right]
+\mathrm{c.c.},
\end{equation}
where $\omega_0=2k_1z_0$ and $\Delta\omega=\frac{4\ln2}{\sigma T}$ defines the frequency resolution.

In the small-amplitude limit $k_0\delta_{z_0}\ll1$, the Bessel functions satisfy
\[
\frac{J_{\pm1}(2k_0\delta_{z_0})}{J_0(2k_0\delta_{z_0})}
\approx
\pm k_0\delta_{z_0}.
\]
This approximation remains valid for $k_0\delta_{z_0}<1$ (corresponding to $\delta_{z_0}\lesssim200~\mathrm{nm}$ for a 1310-nm laser). Retaining only first-order modulation terms in Eq.~\eqref{eq:photocurrent_FT} gives
\begin{equation}
\begin{aligned}
F_j(\omega,\omega_0)\approx
F_0(\omega_0)\Big[
-k_0\delta_{z_0}e^{-i\varphi_{z_0}^j}
e^{-\ln2\frac{(\omega-\omega_0+\omega_m)^2}{(\Delta\omega)^2}} +
e^{-\ln2\frac{(\omega-\omega_0)^2}{(\Delta\omega)^2}} +
k_0\delta_{z_0}e^{i\varphi_{z_0}^j}
e^{-\ln2\frac{(\omega-\omega_0-\omega_m)^2}{(\Delta\omega)^2}}
\Big]
+\mathrm{c.c.},
\end{aligned}
\label{eq:spectraldomain}
\end{equation}
where the three terms correspond to the left sideband, main lobe, and right sideband, respectively, and
\[
F_0(\omega_0)=
\frac{r_{z_0}}{2}
\left(\sqrt{\frac{\pi}{\ln2}}\frac{\sigma T}{2}\right)
J_0(2k_0\delta_{z_0})
e^{i2k_0z_0}.
\]

\subsection{Demodulation algorithm}

In practice, the spectral signal near $\omega_0$ consists of a static component $F_j(\omega_0,\omega_0)$ and two vibration-induced sidebands originating from the adjacent depths $(z_0-z_m)$ and $(z_0+z_m)$, where $z_m=\frac{\omega_m}{2k_1}$.
The superposed signal can therefore be written as
\begin{equation}
\mathcal{F}_j(\omega,\omega_0)
\approx
F_j(\omega,\omega_0-\omega_m)
+
F_j(\omega,\omega_0)
+
F_j(\omega,\omega_0+\omega_m).
\end{equation}
Evaluating the signal at $\omega_0$ gives
\begin{equation}
\begin{aligned}
\mathcal{F}_j(\omega_0,\omega_0)\approx
&
F_0(\omega_0-\omega_m)
k_0\delta_{z_0-z_m}e^{i\varphi_{z_0-z_m}^j}
+
F_0(\omega_0) \\
&
-
F_0(\omega_0+\omega_m)
k_0\delta_{z_0+z_m}e^{-i\varphi_{z_0+z_m}^j}.
\end{aligned}
\end{equation}

Recovery of the vibration amplitude and phase at $(z_0-z_m)$ requires at least \emph{three} independent acquisitions.
Let the acquisition start times be
\begin{equation}\label{eq:time_pts}
t_1=0,\quad
t_2=\frac{2N\pi+2\pi/3}{\omega_m},\quad
t_3=\frac{4N\pi+4\pi/3}{\omega_m},
\end{equation}
where the integer $N$ is chosen such that $t_2-t_1>T$.
Averaging the three measurements gives the static component
\[
F_0(\omega_0)=\frac{1}{3}\sum_{j=1}^{3}\mathcal{F}_j(\omega_0,\omega_0).
\]
Subtracting the mean removes the static contribution,
\begin{equation}
\begin{aligned}
\mathcal{F}_j(\omega_0,\omega_0)
-
\frac{1}{3}\sum_{j=1}^{3}\mathcal{F}_j(\omega_0,\omega_0)
=&
F_0(\omega_0-\omega_m)
k_0\delta_{z_0-z_m}
e^{i(\varphi_{z_0-z_m}+(j-1)\frac{2\pi}{3})}\\
&-
F_0(\omega_0+\omega_m)
k_0\delta_{z_0+z_m}
e^{-i(\varphi_{z_0+z_m}+(j-1)\frac{2\pi}{3})}.
\end{aligned}
\end{equation}
Normalizing by the averaged signal at $(\omega_0-\omega_m)$ defines
\[
G_j=
\frac{
\mathcal{F}_j(\omega_0,\omega_0)
-\frac{1}{3}\sum_{j=1}^{3}\mathcal{F}_j(\omega_0,\omega_0)
}{
\frac{1}{3}\sum_{j=1}^{3}\mathcal{F}_j(\omega_0-\omega_m,\omega_0-\omega_m)
}.
\]
The vibration amplitude and phase at $(z_0-z_m)$ are then obtained as
\begin{equation}
\frac{1}{k_0}
\frac{G_2-G_1e^{-i2\pi/3}}
{e^{i2\pi/3}-e^{-i2\pi/3}}
=
\delta_{z_0-z_m}e^{i\varphi_{z_0-z_m}}.
\label{eq:solution}
\end{equation}
This expression isolates the complex vibration component at depth $z_0-z_m$ while suppressing the conjugate contribution from $z_0+z_m$. Consequently, the proposed demodulation scheme remains valid for arbitrary $\omega_m$, irrespective of spectral aliasing~\cite{RN39}, providing a general reconstruction framework for ultrafast OCE.

For low-frequency excitation, the sidebands merge with the main lobe, and the vibration information is encoded in the phase of the interference signal $\phi$. In this regime, the vibration signal can be recovered directly from the phases of the three acquisitions by fitting $\phi_1$, $\phi_2$, and $\phi_3$ with a sinusoidal function with a constant phase offset (see SI, Note 1).

\subsection{Synchronization for Ultrafast OCE}

Ultrafast OCE operates in the M--B scan mode to acquire a two-dimensional wave field. Let $N_b$ denote the number of A-lines in each B-scan. For the $n$-th ($n = 0, 1, 2, \cdots, N_b - 1$) lateral position within the B-scan, the acquisition time points for the three measurements are shifted by $n/f_a$:
\begin{equation}\label{eq:time_pts2}
t^n_1= \frac{n}{f_a},\quad
t^n_2=\frac{n}{f_a} + \frac{2N\pi+2\pi/3}{\omega_m},\quad
t^n_3=\frac{n}{f_a} + \frac{4N\pi+4\pi/3}{\omega_m}.
\end{equation}

According to Eq.~\eqref{eq:solution}, the vibration signal recovered from the three acquisitions at $t^n_1$, $t^n_2$, and $t^n_3$ can be expressed as
\begin{equation}\label{eq:solution2}
\frac{1}{k_0}
\frac{G^n_2-G^n_1e^{-i2\pi/3}}
{e^{i2\pi/3}-e^{-i2\pi/3}}
=
\delta_{z^n_0-z_m}e^{i(\varphi_{z^n_0-z_m} + 2n\pi f_m/f_a)}.
\end{equation}

Here, $\delta_{z^n_0-z_m}$ and $\varphi_{z^n_0-z_m}$ denote the vibration amplitude and initial phase of the scatterer at depth $z^n_0=\frac{\omega^n_0}{2k_1}$ in the $n$-th A-line. The recovered phase contains an additional asynchronous phase shift of $2n\pi f_m/f_a$, which can be removed by multiplying the left-hand side of Eq.~\eqref{eq:solution2} by the correction factor $e^{-i2n\pi f_m/f_a}$.

\subsection{Experimental setup}

Our experimental setup is based on a home-built swept-source optical coherence elastography (OCE) system.
The system employs a HSL-20 (Santec) wavelength-swept laser with a central wavelength of $1310$~nm, a 3-dB bandwidth of $105$~nm, and an A-line rate of $100$~kHz, providing an axial resolution of $15~\mu$m.
The optical beam is scanned transversely by a pair of galvanometer mirrors and focused by a wide-aperture scan lens (LSM54-1310, Thorlabs), providing a long working distance of $64$~mm and a lateral resolution of $30~\mu$m.
The average optical power delivered to the sample is $13.4$~mW.

A fiber Bragg grating (FBG) and a photodiode (PD) generate a pulse signal synchronized with each wavelength sweep, serving as an optical clock for synchronizing the PZT excitation, OCT beam scanning, and data acquisition.
The interferometric signal from a dual-balanced detector (PDB435C, Thorlabs) is digitized by a high-speed data acquisition board (ATS9371, AlazarTech) at a sampling rate of $300$~MHz, with $1328$ data points acquired per A-line.
An input/output (I/O) board (USB-6353, National Instruments) generates the analog waveforms for the galvanometer scanners while simultaneously recording the stimulus waveforms applied to the PZT.
The excitation signals are generated by a function generator (DG4062, RIGOL), amplified, and applied to the PZT.

For frequencies below $100$~kHz, a custom mechanical actuator composed of a PZT (PA4FLW, Thorlabs) and a 3D-printed prism-shaped tip with a line contact length of $5$~mm is used to reduce wave attenuation and suppress supershear surface waves.
For frequencies above $100$~kHz, a PZT (PA4CEW, Thorlabs) is placed in direct contact with the sample over a contact area of $2$~mm $\times$ $2$~mm to provide sufficient excitation force.
The corresponding amplifiers are LYAP-150BDS (Longyi) for low-frequency operation and ATA-8152 (Aigtek) for high-frequency operation.

In the continuous B-scan protocol, the OCT beam is translated laterally across the sample, acquiring one A-line at each transverse position to form a B-scan with $N_b$ A-lines.
After each B-scan, the beam returns to the starting position and repeats the scan.
Typically, $N_b$ ranges from $100$ to $300$.
For $N_b=150$, one B-scan requires approximately $1.5$~ms, and three B-scans are sufficient to complete one OCE measurement, resulting in a total acquisition time of $4.5$~ms.

\subsection{Displacement field analysis}

All algorithms were implemented in MATLAB R2021b (MathWorks, Inc.).
The raw B-scan data were processed using a standard swept-source phase-stabilization algorithm to obtain complex-valued OCT tomograms.
A demodulation algorithm was then applied to extract the full-field asynchronous amplitude and phase from three consecutive B-scans.
A synchronization algorithm was subsequently applied to correct the phase differences across lateral positions, thereby reconstructing the synchronized displacement field.

After reconstructing the displacement field, one- or two-dimensional Fourier transforms were applied to the surface-wave field to convert the data from the spatial domain to the wavenumber (\(k\)) domain.
The surface-wave wavenumber \(k\) was determined from the spectral peak, and the phase velocity was calculated as \(v = 2\pi f_m / k\).

When \(f_m<f_a\), a demodulation algorithm based on the interference phase can also be used.
This approach produces results identical to those of the complex demodulation algorithm while providing a convenient means of validating both the data and the implementation.
In addition, motion artifacts caused by linear sample motion were corrected using data acquired from a fourth B-scan.

\subsection{Preparation of materials}

An acrylic plate with a thickness of $0.75$~mm and an area of $5$~mm $\times$ $5$~mm was directly bonded to the surface of the PZT.
The PEGDA sample had a diameter of $60$~mm and a height of $12$~mm.
PEGDA was mixed with deionized water at a ratio of $1:9$.
After thorough mixing, ammonium persulfate was added as an initiator at $1\%$ of the total weight, and the solution was stirred until completely dissolved.
The mixture was then poured into a mold and cured at $90\,^\circ\mathrm{C}$ for $30$~min.
The mass density was assumed to be \(\rho \approx 1000~\mathrm{kg/m^3}\).

\subsection{Animal experiments}

All animals were anesthetized with a single dose of $3\%$ sodium pentobarbital.

For the vascular dynamic measurement experiments, adult rats were used.
After anesthesia, the carotid arteries were surgically exposed.
The animals were fixed on a translation stage, and the carotid artery was immobilized using forceps.
The stage was adjusted to gently bring the carotid artery into contact with the probe and within the OCT imaging range before data acquisition.

For the three-dimensional OCE experiments, SPF-grade adult male New Zealand White rabbits were used.
After anesthesia, the rabbits were secured on a translation stage.
The eyelids were retracted using an eyelid speculum, and topical anesthesia was administered.
The stage was then adjusted to gently bring the probe into contact with the sclera within the OCT imaging range, followed by data acquisition.

\subsection{Human participants}

This study was approved by the Biomedical Ethics Committee of Peking University (Approval No.~IRB00001052-25213), and written informed consent was obtained from the participant before the experiment.

The volar pad of the middle phalanx of the left middle finger of a 23-year-old male participant was examined.
Measurements were performed immediately after disinfection of the imaging site.

\subsection{Statistics and reproducibility}

All statistical analyses were performed using Microsoft Excel (Microsoft Inc.) and MATLAB R2021b (MathWorks, Inc.).
All displacement-field measurements reported in the main text and the Supplementary Information were independently repeated at least three times with consistent results.

\section{Results}

\subsection{Ultrafast OCE with phase-shifted scanning}

\begin{figure}[b!]
\centering
\includegraphics[width=1\textwidth]{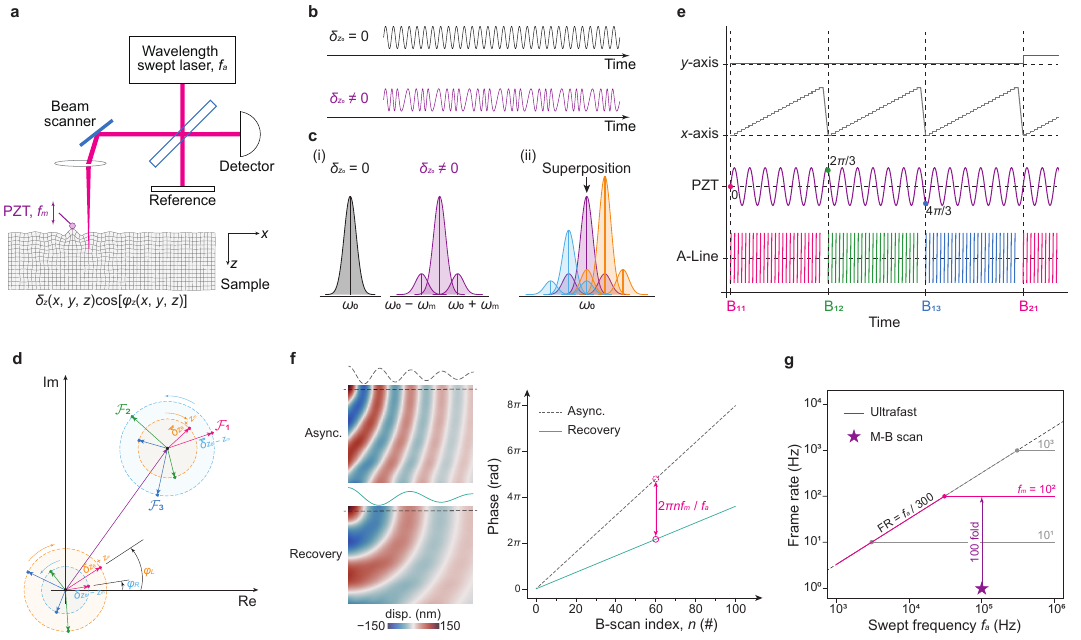}
\caption{\textbf{Principle of ultrafast OCE.}
(a) Schematic of ultrafast OCE. Shear waves are excited by a PZT driven at frequency $f_m$. Full-field shear waves are reconstructed from the vibration amplitude $\delta_z(x,y,z)$ and phase $\varphi_z(x,y,z)$ measured by OCT.
(b) Time-domain interference signals from a stationary scatterer ($\delta_{z_0}=0$) and a harmonically vibrating scatterer ($\delta_{z_0}\neq0$).
(c) (i) Fourier transforms (FTs) of the interference signals. Harmonic vibration generates two side lobes in addition to the static main lobe. As the excitation frequency $\omega_m=2\pi f_m$ increases, the side lobes separate from the main lobe. (ii) At each pixel, the detected signal is the superposition of the static main lobe and two side lobes originating from adjacent scatterers.
(d) Principle of phase-shifted demodulation. The static main lobe is first removed by shifting the rotating phase circles to the origin. Three equally phase-shifted acquisitions are then used to recover the vibration amplitude $\delta_{z_0}$ and phase $\varphi_{z_0}$.
(e) Scanning scheme of ultrafast OCE. Two-dimensional wave fields are acquired using three B-scans initiated at excitation phases $0$, $2\pi/3$, and $4\pi/3$.
(f) Wave field reconstructed by pixel-wise demodulation before phase correction. The horizontal phase gradient (dashed line) arises from the additional phase shift $2\pi n f_m/f_a$ introduced by asynchronous B-scan acquisition. Removing this phase shift recovers the correct full wave field.
(g) Imaging rate of ultrafast OCE as a function of excitation frequency $f_m$ for a swept-source laser operating at $f_a$. For a representative case with $f_a=100$~kHz and $f_m=100$~Hz, ultrafast OCE achieves an imaging rate approximately two orders of magnitude higher than conventional OCE.}
\label{fig:1}
\end{figure}

Ultrafast OCE reconstructs the complete shear-wave vibration field from only three phase-shifted B-scans, enabling full-field elastography on a conventional swept-source OCT system (Fig.~\ref{fig:1}a). The interferometric OCT signal provides the vibration amplitude $\delta_z(x,y,z)$ and phase $\varphi_z(x,y,z)$ required for elastogram reconstruction.

A scatterer vibrating harmonically at depth $z_0$ periodically phase-modulates the OCT interference signal (Fig.~\ref{fig:1}b), producing two vibration-induced side lobes in addition to the static main lobe after Fourier transformation (Fig.~\ref{fig:1}c). As demonstrated previously~\cite{RN39}, the amplitudes and phases of the two side lobes uniquely determine the vibration amplitude and phase, enabling quantitative measurements from acoustic to ultrasonic frequencies.

For distributed scatterers, the detected signal at each pixel is the superposition of a static main lobe and two side lobes originating from adjacent depths. Consequently, the three components cannot be separated from a single acquisition. We resolve this ambiguity by acquiring three datasets with excitation phases separated by $2\pi/3$ (see Materials and Methods). In the complex plane, the static main lobe remains stationary whereas the two side lobes rotate in opposite directions, allowing the three acquisitions to uniquely separate the three components and recover the vibration amplitude and phase at every voxel (Fig.~\ref{fig:1}d). This phase-shifted demodulation requires only three acquisitions, the theoretical minimum for complete wavefield reconstruction.

Below approximately $100$ kHz, the side lobes merge with the main lobe, and the vibration is instead encoded in the phase of the reconstructed OCT signal. In this regime, the displacement is directly obtained from the phase change, $\Delta\phi\lambda_0/(4\pi)$~\cite{RN60}.

To acquire two-dimensional wave fields, three B-scans are synchronized with excitation phases of $0$, $2\pi/3$, and $4\pi/3$, respectively (Fig.~\ref{fig:1}e). The beam then scans laterally while repeating the three phase-shifted acquisitions at each position. Because neighboring A-lines are acquired sequentially, an additional deterministic phase shift of $2\pi n f_m/f_a$ is introduced along the scanning direction (Fig.~\ref{fig:1}f). This deterministic phase shift is removed analytically to recover the correct full-field wave pattern.

The achievable imaging rate is determined by the swept-source speed $f_a$ and excitation frequency $f_m$ (Fig.~\ref{fig:1}g). When $f_a \le 3N_bf_m$, the imaging rate equals $f_a/(3N_b)$. Otherwise, the imaging rate reaches the excitation frequency $f_m$ (see SI, Note~2). For a representative system with $N_b=100$, $f_a=100$ kHz, and $f_m=100$ Hz, ultrafast OCE achieves an imaging rate of $100$ Hz, representing a 100-fold increase over conventional M--B scan-based OCE.

\subsection{Ultrafast OCE across acoustic and ultrasonic regimes}

Ultrafast OCE was implemented on a swept-source OCT system operating at an A-line rate of $f_a=100$~kHz. The system employed a $1310$-nm swept-source laser with a $105$-nm bandwidth, providing an axial resolution of approximately $15~\mu$m (see Materials and Methods). This platform enables ultrafast OCE over excitation frequencies spanning both the acoustic and ultrasonic regimes.

\begin{figure}[htb!]
\centering
\includegraphics[width=1.\textwidth]{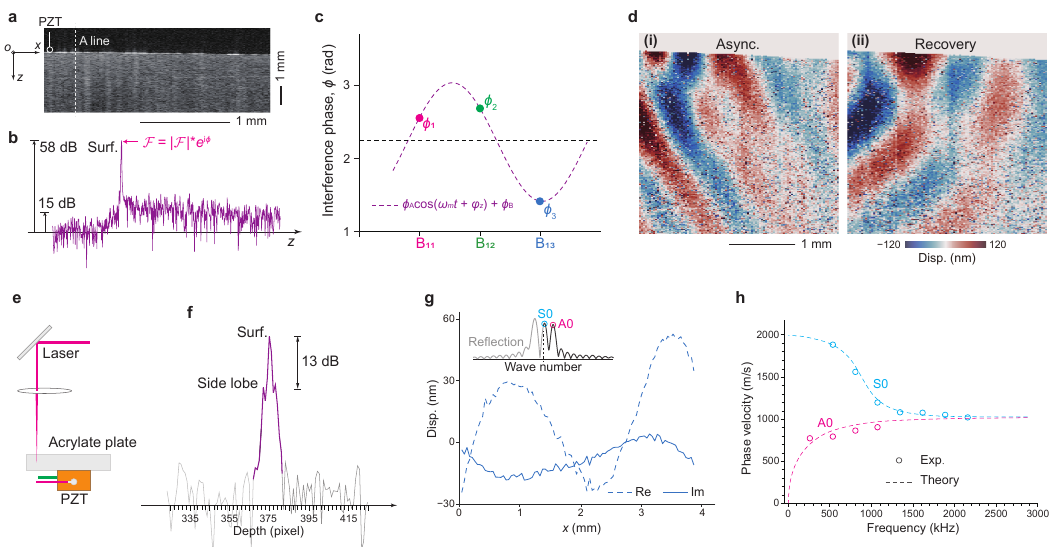}
\caption{\textbf{Experimental demonstration of ultrafast OCE.}
(a) OCT image of the gel phantom used to validate ultrafast OCE in the acoustic regime ($4$~kHz).
(b) Representative OCT A-line profile showing the sample surface and internal scatterers.
(c) Interference phases of the surface signal extracted from three phase-shifted B-scans. A biased cosine fit determines the vibration amplitude, $\delta_z=\phi_A\lambda_0/(4\pi)$, and phase, $\varphi_z$.
(d) Full-field shear-wave displacement maps before (i) and after (ii) correction of the deterministic phase drift introduced by asynchronous scanning.
(e) Experimental setup for validating ultrafast OCE in the ultrasonic regime using an acrylate plate excited over frequencies from $0.2$ to $2.2$~MHz.
(f) Representative OCT spectrum of the sample surface at $f_m=0.81$~MHz, showing the static main lobe and two vibration-induced side lobes separated by three pixels.
(g) Surface displacement reconstructed by ultrafast OCE. Inset: Fourier transform of the displacement field identifying the fundamental antisymmetric (A$_0$) and symmetric (S$_0$) Lamb modes.
(h) Measured Lamb-wave dispersion curves compared with the elastic Lamb-wave model, demonstrating excellent agreement over the entire frequency range.}
\label{fig:2}
\end{figure}

We first validated ultrafast OCE in the acoustic regime using a soft gel phantom excited by a PZT at $4$~kHz (Fig.~\ref{fig:2}a). The representative OCT depth profile in Fig.~\ref{fig:2}b exhibited optical SNRs of approximately $58$ and $15$~dB for the surface and internal scatterers, respectively, providing sufficient signal quality for vibration reconstruction. Three phase-shifted B-scans were then acquired using the ultrafast scanning scheme (Fig.~\ref{fig:1}e). After correcting the deterministic phase drift introduced by asynchronous scanning, the full-field shear-wave displacement was reconstructed (Figs.~\ref{fig:2}c,d). The reconstructed wave field agrees closely with conventional M--B scan-based OCE measurements (see SI, Fig.~S1), while reducing the acquisition time from approximately $1$~s to only $3$~ms.

We next extended ultrafast OCE to the ultrasonic regime using an acrylate plate bonded to a PZT actuator and excited over the frequency range $0.2$--$2.2$~MHz (Fig.~\ref{fig:2}e). Owing to the optical transparency of the sample, the OCT signal was dominated by the sample surface reflection, enabling direct reconstruction of the surface displacement field after phase-shifted demodulation (Figs.~\ref{fig:2}f, g). The reconstructed wave field comprises multiple laterally propagating Lamb modes. Fourier analysis identified the fundamental antisymmetric (A$_0$) and symmetric (S$_0$) modes, from which the phase velocities were extracted (Fig.~\ref{fig:2}g, inset). The measured dispersion curves agree closely with the elastic Lamb-wave model across the entire frequency range (Fig.~\ref{fig:2}h), validating ultrafast OCE in the ultrasonic regime. Under these conditions, ultrafast OCE achieves an imaging rate of $\sim333$~Hz, demonstrating high-speed elastography in the ultrasonic regime.

\subsection{Ultrafast OCE enhances vibration detection sensitivity}

\begin{figure}[t!]
\centering
\includegraphics[width=1.\textwidth]{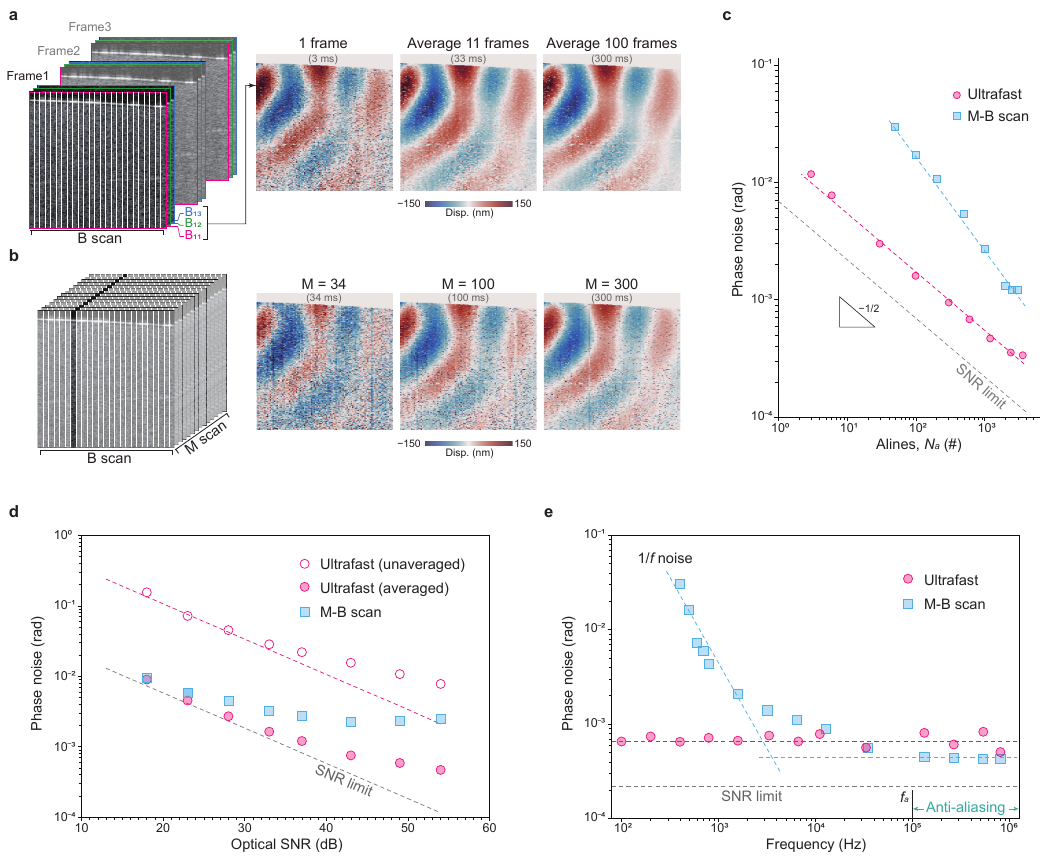}
\caption{\textbf{Sensitivity of ultrafast OCE.}
(a) Full-field shear-wave displacement in a gel phantom measured by ultrafast OCE. Each frame is reconstructed from three phase-shifted B-scans, and multiple frames are averaged to improve the signal-to-noise ratio (SNR).
(b) Full-field shear-wave displacement measured by conventional M--B scan-based OCE. Results for different M-scan lengths.
(c) Comparison of phase stability between ultrafast OCE and conventional M--B scan-based OCE.
(d) Phase stability as a function of optical SNR.
(e) Phase stability over excitation frequencies spanning the acoustic and ultrasonic regimes.}
\label{fig:3}
\end{figure}

We next compare the vibration detection sensitivity of ultrafast OCE with that of conventional M--B scan-based OCE. Full-field shear waves excited at approximately $3$~kHz were imaged in a gel phantom using both methods (Figs.~\ref{fig:3}a,b). To ensure a fair comparison, the two approaches were implemented with the same lateral sampling density of $100$ points per B-scan.

Ultrafast OCE reconstructs a full-field shear-wave image from only three B-scans, requiring only $3$~ms of acquisition time. In contrast, conventional M--B scan-based OCE requires at least $34$ A-lines per lateral position to sample one excitation cycle, resulting in a minimum acquisition time of approximately $34$~ms under identical imaging conditions. Despite this order-of-magnitude reduction in acquisition time, ultrafast OCE reconstructs a visibly cleaner wave field (Figs.~\ref{fig:3}a,b). Averaging 11 ultrafast frames to match the total acquisition time of the M--B scan further improves the reconstructed wave field quality. These results demonstrate that ultrafast OCE improves vibration detection sensitivity while simultaneously enabling much faster full-field imaging.

To quantify the sensitivity improvement, we measured the phase stability using a mirror-like sample with controlled optical SNR driven by a $1$~kHz PZT (see SI, Fig.~S2). Phase stability was evaluated as the standard deviation of 100 repeated phase measurements. Figure~\ref{fig:3}c summarizes the phase stability as a function of the total number of acquired A-lines, while keeping the total acquisition time identical for both methods. The optical SNR was fixed at $48$~dB. Ultrafast OCE requires only three A-lines for wavefield reconstruction, and the phase stability improves with averaging following the expected $N_a^{-1/2}$ scaling. In contrast, the M--B scan exhibits substantially poorer phase stability for short M-scans and approaches the ultrafast performance only when the M-scan length exceeds approximately $2000$ A-lines. Consequently, ultrafast OCE consistently achieves superior phase stability under the same acquisition time.

We next examine the phase stability as a function of optical SNR (Fig.~\ref{fig:3}d). At low optical SNR, both ultrafast OCE and conventional M--B scan-based OCE follow the theoretical shot-noise limit (gray dashed lines). As the optical SNR exceeds approximately $30$~dB, however, the measured phase stability deviates from the theoretical prediction because additional noise sources, including environmental fluctuations and electronic noise, become dominant. Despite these non-ideal noise contributions, ultrafast OCE consistently exhibits superior phase stability, demonstrating greater robustness for high-SNR measurements.

We finally evaluate the vibration sensitivity over excitation frequencies spanning the acoustic and ultrasonic regimes (Fig.~\ref{fig:3}e). While both methods exhibit comparable phase stability above approximately $1$~kHz, ultrafast OCE maintains high sensitivity at low frequencies, eliminating the strong frequency dependence observed for conventional M--B scan-based OCE. This advantage arises from the fundamentally different acquisition strategies of the two methods. Ultrafast OCE samples the vibration at the minimum rate required for phase-shifted demodulation ($f_m/3$), while extending the effective integration time through cross-triplet averaging (see SI, Fig.~S2). In contrast, conventional M--B scan-based OCE samples each M-scan at the full A-line rate ($f_a=100$~kHz), resulting in a short effective averaging time and greater susceptibility to low-frequency electronic noise. Consequently, ultrafast OCE is inherently more robust to $1/f$ noise, providing consistently high vibration detection sensitivity from acoustic to ultrasonic frequencies (see SI, Note~3).

\subsection{Ultrafast OCE captures rapidly evolving tissue mechanics}

\begin{figure}[t]
\centering
\includegraphics[width=1.\textwidth]{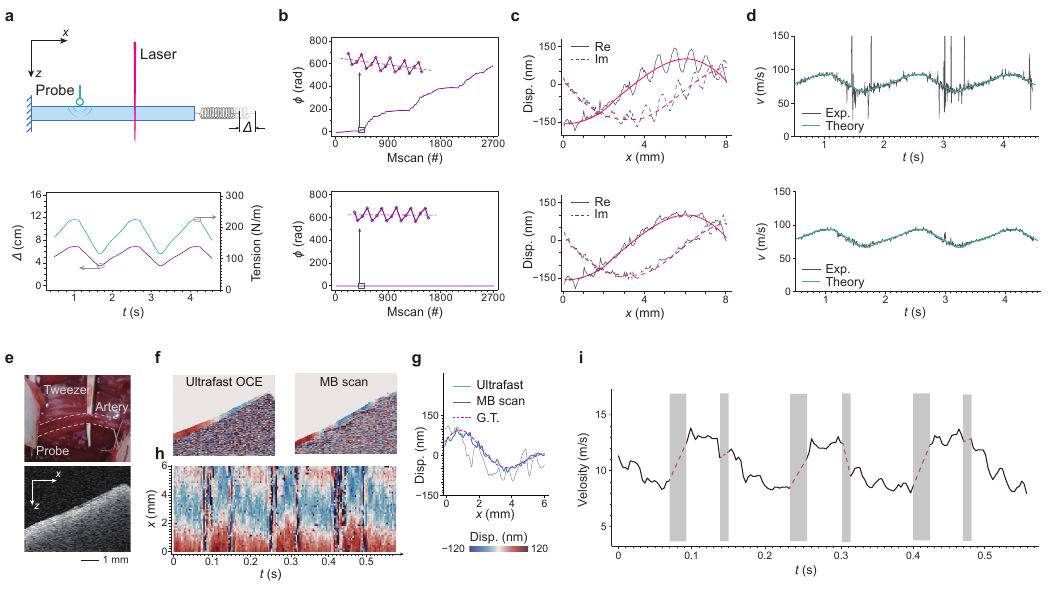}
\caption{\textbf{Ultrafast OCE for probing dynamic changes in mechanical properties.}
(a) Experimental setup for dynamic thin-film measurements.
(b) Interference phase before and after motion correction.
(c) Reconstructed wave fields before and after motion correction.
(d) Dynamic wave-speed measurements of the thin film.
(e) Photograph and OCT image of the rat carotid artery.
(f) Representative wave fields measured by ultrafast and conventional M--B scan-based OCE.
(g) Representative surface displacement profiles.
(h) Temporal evolution of the surface displacement profiles over approximately three cardiac cycles.
(i) Temporal evolution of arterial wave speed.}
\label{fig:4}
\end{figure}

With frame rates reaching hundreds of hertz, ultrafast OCE enables real-time imaging of rapidly evolving tissue mechanics. To demonstrate this capability, we first measured a thin elastic film subjected to periodic tensile loading (Fig.~\ref{fig:4}a). The film was repeatedly stretched by a reciprocating motor, producing cyclic variations in tension with a period of approximately $1.6$~s, comparable to the time required to acquire a single frame using conventional M--B scan-based OCE. Elastic waves were generated using a contact probe operating at $f_m=6.6$~kHz, while $2700$ consecutive B-scans were acquired throughout the loading cycle.

Motion of the stretched film produces a slow drift in the interference phase, as illustrated by the representative scatterer in Fig.~\ref{fig:4}b. After motion correction (see SI Note 5), the phase trace becomes nearly stationary, allowing every three consecutive B-scans to be demodulated into one vibration frame with an effective imaging rate of approximately $180$~Hz. Figure~\ref{fig:4}c shows a representative reconstructed wave field after motion correction, demonstrating effective suppression of motion artifacts. Fourier analysis of the reconstructed wave field yields the instantaneous wave speed (Fig.~\ref{fig:4}d), with motion correction substantially improving the precision of wave-speed estimation. The dynamic measurements were independently validated using both experiment and theory. The measured wave speeds in the relaxed and stretched states agreed closely with those obtained using conventional M--B scan-based OCE (see SI, Fig.~S3). Moreover, the measured temporal evolution closely followed the acoustoelastic prediction (teal curve in Fig.~\ref{fig:4}d), confirming the quantitative accuracy of ultrafast OCE during dynamic loading.

We next applied ultrafast OCE to measure pulsatile changes in arterial mechanics \emph{in vivo} using the carotid artery of a rat (Fig.~\ref{fig:4}e). The rat heart rate is approximately $350$ beats per minute (bpm), producing rapid cyclic variations in arterial stiffness that are difficult to capture using conventional M--B scan-based OCE. After surgically exposing the carotid artery, elastic waves were excited with a contact probe operating at $f_m=3$~kHz, corresponding to an effective imaging rate of approximately $300$~Hz. Ultrafast OCE was performed from a longitudinal imaging plane, and a continuous acquisition lasting $0.56$~s captured approximately three cardiac cycles.

Figure~\ref{fig:4}f compares representative wave fields acquired using ultrafast OCE and conventional M--B scan-based OCE. Owing to its high imaging speed, ultrafast OCE reconstructs continuous wave patterns that clearly delineate the arterial wall, whereas conventional M--B scan-based OCE fails to recover a complete wave field owing to motion-induced distortion. Surface displacement profiles extracted from successive wave fields are shown in Figs.~\ref{fig:4}g,h. Wavefield reconstruction fails only during two short intervals within each cardiac cycle, corresponding to pulse-wave propagation and the dicrotic phase, when rapid arterial motion exceeds the motion-correction limit (see Materials and Methods). Outside these intervals, the recovered displacement profiles remain highly sinusoidal, enabling reliable wave-speed estimation throughout most of the cardiac cycle.

The arterial wave speed varies periodically throughout the cardiac cycle (Fig.~\ref{fig:4}i), reaching a maximum of $13.2\pm0.5$~m/s during systole and a minimum of $8.4\pm0.1$~m/s during diastole. These variations reflect the nonlinear elasticity of collagen-rich arterial tissue, whose stiffness increases with blood pressure~\cite{RN1714}. The measured $\sim61\%$ change in wave speed is substantially larger than that reported for human arteries ($\sim20\%$~\cite{RN1498}), likely reflecting the larger blood-pressure fluctuations in rats ($147.8\pm5.6$~mmHg systolic and $98.3\pm5.2$~mmHg diastolic~\cite{RN1713}). These results demonstrate that ultrafast OCE enables direct imaging of rapidly evolving arterial stiffness \emph{in vivo}.

\subsection{Ultrafast OCE enables in vivo volumetric imaging}

By extending the lateral scan to two dimensions, ultrafast OCE enables sub-second volumetric \emph{in vivo} imaging. To demonstrate this capability, we performed volumetric imaging of human finger skin and the anterior eye globe of a rabbit. In both experiments, shear waves were generated using a contact probe driven by a PZT actuator.

\begin{figure}[ht!]
\centering
\includegraphics[width=1.\textwidth]{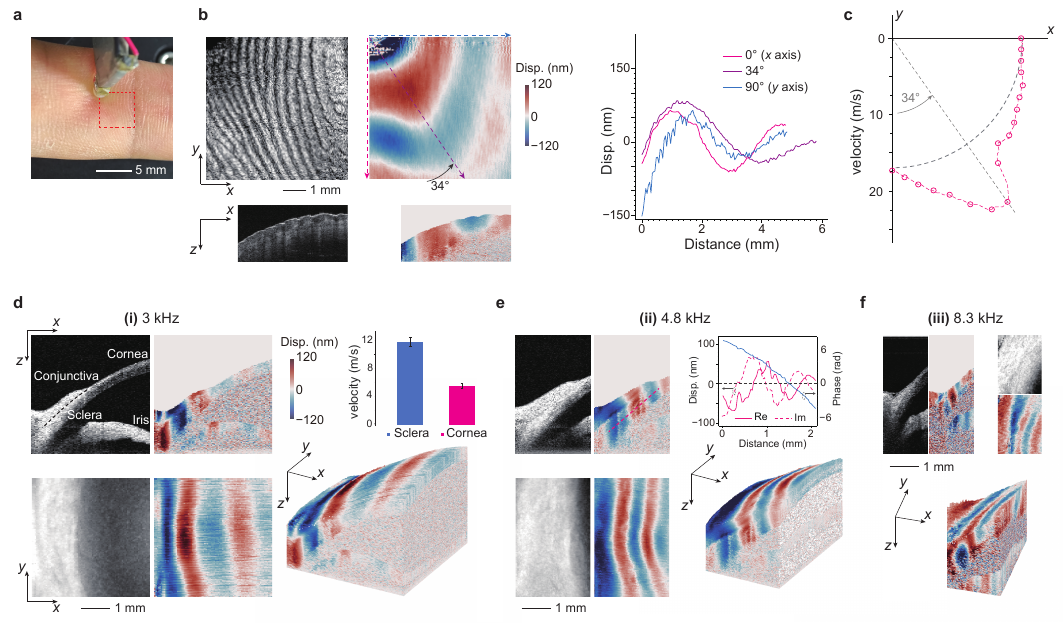}
\caption{\textbf{Ultrafast OCE for volumetric shear wave imaging.}
(a) Volumetric imaging of human finger skin in vivo.
(b) Structural OCT image and the corresponding shear-wave field. Wave profiles extracted along three propagation directions ($0^\circ$, $34^\circ$, and $90^\circ$).
(c) Direction-dependent shear-wave velocity, reaching a maximum parallel to the fingerprint ridges.
(d)--(f) Volumetric shear-wave imaging of the anterior eye globe in a rabbit. The reconstructed wave fields reveal pronounced mechanical heterogeneity, including layered interfaces within the sclera and a gradual stiffness transition from the sclera to the cornea.}
\label{fig:5}
\end{figure}

We first performed volumetric imaging of the middle phalanx of a 23-year-old volunteer (Fig.~\ref{fig:5}a; see also Materials and Methods). This site exhibits well-defined fingerprint ridges and multilayer tissue architecture, providing a representative model for evaluating volumetric imaging performance. A $150\times150$ scan grid was used to acquire a volumetric dataset in $0.9$~s. Figure~\ref{fig:5}b shows the corresponding structural OCT image and reconstructed shear-wave field of the skin. The OCT image resolves the surface ridge--groove pattern, while the shear-wave field captures concentric wave propagation throughout the imaging volume.

The reconstructed wave field reveals pronounced directional anisotropy, with wavelengths significantly longer along a preferred propagation direction (Fig.~\ref{fig:5}b). Quantitative analysis (Fig.~\ref{fig:5}c) shows that the shear-wave velocity reaches a maximum of $25.7$~m/s at approximately $34^\circ$, compared with approximately $17$~m/s along the horizontal and vertical scan directions, corresponding to an anisotropy ratio of about $1.5$. This anisotropic behavior is consistent with the organization of fingerprint ridges and is likely associated with the corrugated surface morphology, which produces different bending stiffnesses parallel and perpendicular to the ridges~\cite{RN1755}. These results demonstrate that volumetric imaging enables quantitative characterization of direction-dependent tissue mechanics beyond the capability of two-dimensional imaging.

We next performed \emph{in vivo} volumetric imaging of the anterior eye globe in a rabbit model to demonstrate the capability of ultrafast OCE in a structurally heterogeneous tissue~\cite{RN469}. Shear waves were generated by placing the actuator on the conjunctiva and exciting it at frequencies of $3$, $4.8$, and $8.3$~kHz. The reconstructed volumetric shear-wave fields (Figs.~\ref{fig:5}d--f) reveal pronounced mechanical heterogeneity throughout the anterior eye, demonstrating the capability of ultrafast OCE for volumetric biomechanical imaging in structurally heterogeneous tissues \emph{in vivo}.

The reconstructed shear-wave fields reveal structural and mechanical features that are not readily apparent in the corresponding structural OCT images. In the $x$--$z$ view, discontinuities in the wave field at $4.8$~kHz clearly delineate the layered interfaces within the sclera (Fig.~\ref{fig:5}e). In the $x$--$y$ view, the shorter wavelengths observed in the conjunctival region indicate its lower stiffness compared with the adjacent sclera and cornea, allowing the conjunctiva, sclera, and cornea to be distinguished (Fig.~\ref{fig:5}d). A gradual decrease in wavelength from the sclera to the cornea is further confirmed by the nonlinear phase variations extracted along the cornea--sclera transition (Fig.~\ref{fig:5}e), indicating a continuous stiffness gradient. Quantitative analysis yields average shear-wave velocities of $11.8\pm0.6$~m/s in the sclera and $5.5\pm0.3$~m/s in the cornea, in good agreement with previous reports that the scleral wave speed is approximately twice that of the cornea~\cite{RN337}.

Together, these results demonstrate that ultrafast OCE enables high-SNR volumetric shear-wave imaging in living tissues on sub-second timescales. The resulting three-dimensional wave fields reveal tissue interfaces, spatial stiffness gradients, and direction-dependent mechanical anisotropy, all of which are difficult to characterize using conventional OCE because of its limited acquisition speed.

\section{Discussion}

Ultrafast OCE overcomes the long-standing acquisition-speed bottleneck of optical coherence elastography, enabling biomechanical measurements that are difficult or impractical with conventional M--B scan-based approaches. Using a standard swept-source OCT system operating at an A-line rate of 100~kHz, the proposed method achieves approximately two orders of magnitude higher frame rates than conventional OCE while maintaining high displacement sensitivity over a broad excitation-frequency range spanning acoustic to ultrasonic regimes. These capabilities enable dynamic measurement of arterial stiffness during the cardiac cycle as well as sub-second volumetric imaging of shear-wave propagation in the cornea and skin.

The enabling innovation is a phase-shifted B--M scanning strategy, in which the interference signal is sampled at three equally spaced vibration phases. This acquisition scheme separates the static and dynamic components of the interference signal in the complex domain, allowing robust reconstruction of the complete harmonic shear-wave field over a broad frequency range. The recovered wave field is directly compatible with full-wave inversion algorithms~\cite{RN872}, providing a foundation for quantitative elastographic imaging with high spatial resolution. Several studies have explored accelerated OCE acquisition. For example, Schmidt \emph{et al.}~\cite{RN1151} proposed an asynchronous B-scan strategy for imaging semi-reverberant shear-wave fields. While effective in specific applications, such approaches generally recover only partial wave field information, whereas the present method reconstructs the full complex-valued shear-wave field required for quantitative elastography.

The present implementation is ultimately limited by the laser sweep rate, with the imaging frame rate given by \(\mathrm{FR}=f_a/(3N_a)\). Although recent developments in FDML swept sources have substantially increased acquisition speeds, conventional OCE implementations typically require low-frequency excitation to maintain synchronization~\cite{RN1742}, limiting spatial resolution. In contrast, ultrafast OCE supports high-frequency excitation using standard swept-source OCT systems, enabling practical volumetric imaging without substantially sacrificing resolution. Integration with next-generation FDML lasers is expected to further increase imaging speed toward the excitation-frequency limit, opening opportunities for ultrahigh-speed biomechanical imaging at even higher spatial resolution.

Beyond improving imaging speed, ultrafast OCE enables two measurement capabilities that are difficult to achieve using conventional OCE. First, the high temporal resolution allows direct visualization of rapidly evolving biomechanical properties, as demonstrated by the measurement of pulsatile arterial stiffness. This capability provides new opportunities for investigating vascular mechanics, mechanobiology, and dynamic physiological processes. Second, sub-second volumetric imaging enables three-dimensional mapping of tissue mechanics in structurally heterogeneous organs, facilitating quantitative characterization of tissue interfaces, mechanical anisotropy, and spatial stiffness variations in complex biological systems.

This study has several limitations. First, although the proposed motion-correction strategy effectively suppresses artifacts arising from slow sample motion, rapid or spatially heterogeneous motion may still reduce reconstruction accuracy. Second, imaging depth and displacement sensitivity remain constrained by the optical signal-to-noise ratio of the OCT system. Third, the current implementation relies on periodic excitation and is therefore less suitable for transient or non-repetitive mechanical events. Future work may address these limitations through improved motion compensation, adaptive acquisition strategies, and integration with advanced light sources and detection schemes.

In summary, ultrafast OCE removes a fundamental acquisition-speed limitation of optical coherence elastography and enables dynamic and volumetric biomechanical imaging in living tissues. By combining ultrafast acquisition with quantitative reconstruction of complex shear-wave fields, the proposed framework expands the scope of OCE from static mechanical characterization toward real-time, three-dimensional biomechanical imaging, providing new opportunities for both fundamental studies of tissue mechanics and future clinical applications.

\begin{backmatter}
\bmsection{Funding}
This work is supported by the National Natural Science Foundation of China (Grant Nos. 12532016 and 12472176), the Beijing Natural Science Foundation (Grant No. F262034), and the Fundamental Research Funds for the Central Universities, Peking University.


\bmsection{Disclosures}
The authors declare no conflicts of interest.

\bmsection{Data availability} Data underlying the results presented in this paper are not publicly available at this time but may be obtained from the authors upon reasonable request.

\bmsection{Supplemental document}
A supplemental document must be called out in the back matter so that a link can be included. For example, “See Supplement 1 for supporting content.” Note that the Supplemental Document must also have a callout in the body of the paper.

\end{backmatter}

\bibliography{sample}

@article{RN39,
   author = {Feng, Xu and Li, Guo-Yang and Yun, Seok-Hyun},
   title = {Ultra-wideband optical coherence elastography from acoustic to ultrasonic frequencies},
   journal = {Nature Communications},
   volume = {14},
   number = {1},
   pages = {4949},
   ISSN = {2041-1723},
   DOI = {10.1038/s41467-023-40625-y},
   url = {https://doi.org/10.1038/s41467-023-40625-y},
   year = {2023},
   type = {Journal Article}
}

@article{RN60,
   author = {Ramier, Antoine and Tavakol, Behrouz and Yun, Seok-Hyun},
   title = {Measuring mechanical wave speed, dispersion, and viscoelastic modulus of the cornea using optical coherence elastography},
   journal = {Optics Express},
   volume = {27},
   number = {12},
   pages = {16635-16649},
   DOI = {10.1364/OE.27.016635},
   url = {https://opg.optica.org/oe/abstract.cfm?URI=oe-27-12-16635},
   year = {2019},
   type = {Journal Article}
}

@article{Kuklinski2019PropertiesOG,
  title={Properties of Generalized Bessel Functions},
  author={Parker Kuklinski and David A. Hague},
  journal={arXiv: General Mathematics},
  year={2019},
  url={https://api.semanticscholar.org/CorpusID:201698162}
}

@article{RN1498,
   author = {Jiang, Yuxuan and Li, Guo-Yang and Hu, Keshuai and Ma, Shiyu and Zheng, Yang and Jiang, Mingwei and Zhang, Zhaoyi and Wang, Xinyu and Cao, Yanping},
   title = {Simultaneous imaging of bidirectional guided waves probes arterial mechanical anisotropy, blood pressure, and stress synchronously},
   journal = {Science Advances},
   volume = {11},
   number = {32},
   pages = {eadv5660},
   DOI = {doi:10.1126/sciadv.adv5660},
   url = {https://www.science.org/doi/abs/10.1126/sciadv.adv5660},
   year = {2025},
   type = {Journal Article}
}

@article{RN1713,
   author = {Naessens, Daphne M. P. and Coolen, Bram F. and de Vos, Judith and VanBavel, Ed and Strijkers, Gustav J. and Bakker, Erik N. T. P.},
   title = {Altered brain fluid management in a rat model of arterial hypertension},
   journal = {Fluids and Barriers of the CNS},
   volume = {17},
   number = {1},
   pages = {41},
   ISSN = {2045-8118},
   DOI = {10.1186/s12987-020-00203-6},
   url = {https://doi.org/10.1186/s12987-020-00203-6},
   year = {2020},
   type = {Journal Article}
}

@article{RN1714,
   author = {Gasser, T. Christian and Ogden, Ray W and Holzapfel, Gerhard A},
   title = {Hyperelastic modelling of arterial layers with distributed collagen fibre orientations},
   journal = {Journal of The Royal Society Interface},
   volume = {3},
   number = {6},
   pages = {15-35},
   ISSN = {1742-5689},
   DOI = {10.1098/rsif.2005.0073},
   url = {https://doi.org/10.1098/rsif.2005.0073},
   year = {2005},
   type = {Journal Article}
}

@article{RN1708,
   author = {Liu, Zeyang and Chen, Guorui and Jo, Min-Seung and Vogel, Viola and Chen, Jun and Rogers, John A. and Li, Song},
   title = {Mechanomedicine},
   journal = {Nature Reviews Bioengineering},
   volume = {4},
   number = {3},
   pages = {216-235},
   ISSN = {2731-6092},
   DOI = {10.1038/s44222-025-00391-6},
   url = {https://doi.org/10.1038/s44222-025-00391-6},
   year = {2026},
   type = {Journal Article}
}

@article{RN1672,
   author = {Wang, Ning and Chien, Shu and Schwartz, Martin A.},
   title = {Mechanomedicine: Present state and future promise},
   journal = {Proceedings of the National Academy of Sciences},
   volume = {122},
   number = {46},
   pages = {e2509566122},
   DOI = {doi:10.1073/pnas.2509566122},
   url = {https://www.pnas.org/doi/abs/10.1073/pnas.2509566122},
   year = {2025},
   type = {Journal Article}
}

@article{RN1514,
   author = {Alisafaei, Farid and Shakiba, Delaram and Hong, Yuan and Ramahdita, Ghiska and Huang, Yuxuan and Iannucci, Leanne E. and Davidson, Matthew D. and Jafari, Mohammad and Qian, Jin and Qu, Chengqing and Ju, David and Flory, Dashiell R. and Huang, Yin-Yuan and Gupta, Prashant and Jiang, Shumeng and Mujahid, Aliza and Singamaneni, Srikanth and Pryse, Kenneth M. and Chao, Pen-hsiu Grace and Burdick, Jason A. and Lake, Spencer P. and Elson, Elliot L. and Huebsch, Nathaniel and Shenoy, Vivek B. and Genin, Guy M.},
   title = {Tension anisotropy drives fibroblast phenotypic transition by self-reinforcing cell–extracellular matrix mechanical feedback},
   journal = {Nature Materials},
   volume = {24},
   number = {6},
   pages = {955-965},
   ISSN = {1476-4660},
   DOI = {10.1038/s41563-025-02162-5},
   url = {https://doi.org/10.1038/s41563-025-02162-5},
   year = {2025},
   type = {Journal Article}
}

@article{RN1612,
   author = {Cheng, Bo and Li, Moxiao and Lin, Min and Guo, Hui and Xu, Feng},
   title = {Mechanobiology across timescales},
   journal = {Nature Reviews Physics},
   volume = {7},
   number = {11},
   pages = {621-644},
   ISSN = {2522-5820},
   DOI = {10.1038/s42254-025-00874-w},
   url = {https://doi.org/10.1038/s42254-025-00874-w},
   year = {2025},
   type = {Journal Article}
}

@article{RN1739,
   author = {Li, Long and Ji, Jing and Ren, He and Gao, Lilan and Li, Xiaona and Li, Ning and Zhang, Songbai and Tang, Kai and Li, Zedong and Ren, Weiyan and Yao, Qing-ping and Huang, Kai and Gong, He and Shao, Yingfeng and Lin, Xianglong and Wang, Xin and Qian, Xiuqing and Song, Jie and Jiang, Yiran and Chen, Hui and Che, Bo and Lü, Dongyuan and Du, Yu and Feng, Fan and Liu, Yanli and Li, Yan and Luo, Meiying and Du, Ruotian and Zhang, Cunyu and Hu, Guanshuo and Ma, Yufei and Wang, Shutong and Yang, Rui and Pu, Fang and Xiang, Bingjie and Zhang, Ming and Shi, Xinghua and Wang, Lizhen and Li, Bo and Kračun, Damir and Chen, Qian and Elsheikh, Ahmed and Liu, Zi-Jun and Liu, Baoyu and Zhao, Chuanrong and Lü, Yonggang and Zeng, Zhu and Li, Zhiyong and Liu, Yiyao and Wang, Guixue and Tan, Wenchang and Zhang, Chunqiu and Zhang, Min and Lou, Jizhong and Tan, Youhua and Deng, Linhong and Long, Mian and Qi, Ying-Xin and Chen, Weiyi and Xu, Feng and Fan, Yubo and Song, Fan},
   title = {Transformative biomechanics and mechanobiology breakthroughs shaping the future of health and medicine},
   journal = {The Innovation},
   volume = {7},
   number = {4},
   ISSN = {2666-6758},
   DOI = {10.1016/j.xinn.2026.101307},
   url = {https://doi.org/10.1016/j.xinn.2026.101307},
   year = {2026},
   type = {Journal Article}
}

@article{RN1364,
   author = {Casar, Jason R. and McLellan, Claire A. and Shi, Cindy and Stiber, Ariel and Lay, Alice and Siefe, Chris and Parakh, Abhinav and Gaerlan, Malaya and Gu, X. Wendy and Goodman, Miriam B. and Dionne, Jennifer A.},
   title = {Upconverting microgauges reveal intraluminal force dynamics in vivo},
   journal = {Nature},
   volume = {637},
   number = {8044},
   pages = {76-83},
   ISSN = {1476-4687},
   DOI = {10.1038/s41586-024-08331-x},
   url = {https://doi.org/10.1038/s41586-024-08331-x},
   year = {2025},
   type = {Journal Article}
}

@article{RN1328,
   author = {Leartprapun, Nichaluk and Zeng, Ziqian and Hajjarian, Zeinab and Bossuyt, Veerle and Nadkarni, Seemantini K.},
   title = {Laser speckle rheological microscopy reveals wideband viscoelastic spectra of biological tissues},
   journal = {Science Advances},
   volume = {10},
   number = {19},
   pages = {eadl1586},
   DOI = {doi:10.1126/sciadv.adl1586},
   url = {https://www.science.org/doi/abs/10.1126/sciadv.adl1586},
   year = {2024},
   type = {Journal Article}
}

@article{RN1693,
   author = {Grasland-Mongrain, Pol and Zorgani, Ali and Nakagawa, Shoma and Bernard, Simon and Paim, Lia Gomes and Fitzharris, Greg and Catheline, Stefan and Cloutier, Guy},
   title = {Ultrafast imaging of cell elasticity with optical microelastography},
   journal = {Proceedings of the National Academy of Sciences},
   volume = {115},
   number = {5},
   pages = {861-866},
   DOI = {doi:10.1073/pnas.1713395115},
   url = {https://www.pnas.org/doi/abs/10.1073/pnas.1713395115},
   year = {2018},
   type = {Journal Article}
}

@article{RN99,
   author = {Kennedy, Brendan F. and Wijesinghe, Philip and Sampson, David D.},
   title = {The emergence of optical elastography in biomedicine},
   journal = {Nature Photonics},
   volume = {11},
   number = {4},
   pages = {215-221},
   ISSN = {1749-4893},
   DOI = {10.1038/nphoton.2017.6},
   url = {https://doi.org/10.1038/nphoton.2017.6},
   year = {2017},
   type = {Journal Article}
}

@article{RN1520,
   author = {Bouvet, Pierre and Bevilacqua, Carlo and Ambekar, Yogeshwari and Antonacci, Giuseppe and Au, Joshua and Caponi, Silvia and Chagnon-Lessard, Sophie and Czarske, Juergen and Dehoux, Thomas and Fioretto, Daniele and Fu, Yujian and Guck, Jochen and Hamann, Thorsten and Heinemann, Dag and Jähnke, Torsten and Jean-Ruel, Hubert and Kabakova, Irina and Koski, Kristie and Koukourakis, Nektarios and Krause, David and La Cavera, Salvatore and Landes, Timm and Li, Jinhao and Mahmodi, Hadi and Margueritat, Jeremie and Mattarelli, Maurizio and Monaghan, Michael and Overby, Darryl R. and Perez-Cota, Fernando and Pontecorvo, Emanuele and Prevedel, Robert and Ruocco, Giancarlo and Sandercock, John and Scarcelli, Giuliano and Scarponi, Filippo and Testi, Claudia and Török, Peter and Vovard, Lucie and Weninger, Wolfgang J. and Yakovlev, Vladislav and Yun, Seok-Hyun and Zhang, Jitao and Palombo, Francesca and Bilenca, Alberto and Elsayad, Kareem},
   title = {Consensus statement on Brillouin light scattering microscopy of biological materials},
   journal = {Nature Photonics},
   volume = {19},
   number = {7},
   pages = {681-691},
   ISSN = {1749-4893},
   DOI = {10.1038/s41566-025-01681-6},
   url = {https://doi.org/10.1038/s41566-025-01681-6},
   year = {2025},
   type = {Journal Article}
}

@article{RN1441,
   author = {Singh, Manmohan and Hepburn, Matt S. and Kennedy, Brendan F. and Larin, Kirill V.},
   title = {Optical coherence elastography},
   journal = {Nature Reviews Methods Primers},
   volume = {5},
   number = {1},
   pages = {39},
   ISSN = {2662-8449},
   DOI = {10.1038/s43586-025-00406-x},
   url = {https://doi.org/10.1038/s43586-025-00406-x},
   year = {2025},
   type = {Journal Article}
}

@article{RN453,
   author = {Liu, Hsiao-Chuan and Zeng, Yushun and Gong, Chen and Chen, Xiaoyu and Kijanka, Piotr and Zhang, Junhang and Genyk, Yuri and Tchelepi, Hisham and Wang, Chonghe and Zhou, Qifa and Zhao, Xuanhe},
   title = {Wearable bioadhesive ultrasound shear wave elastography},
   journal = {Science Advances},
   volume = {10},
   number = {6},
   pages = {eadk8426},
   DOI = {doi:10.1126/sciadv.adk8426},
   url = {https://www.science.org/doi/abs/10.1126/sciadv.adk8426},
   year = {2024},
   type = {Journal Article}
}

@article{RN1690,
   author = {Sack, Ingolf},
   title = {Magnetic resonance elastography from fundamental soft-tissue mechanics to diagnostic imaging},
   journal = {Nature Reviews Physics},
   volume = {5},
   number = {1},
   pages = {25-42},
   ISSN = {2522-5820},
   DOI = {10.1038/s42254-022-00543-2},
   url = {https://doi.org/10.1038/s42254-022-00543-2},
   year = {2023},
   type = {Journal Article}
}

@article{RN884,
   author = {Singh, Manmohan and Zvietcovich, Fernando and Zevallos-Delgado, Christian and Ambekar, Yogeshwari S. and Aglyamov, Salavat R. and Larin, Kirill V.},
   title = {Whole embryo biomechanics with reverberant optical coherence elastography},
   journal = {Optica},
   volume = {11},
   number = {5},
   pages = {686-692},
   DOI = {10.1364/OPTICA.521367},
   url = {https://opg.optica.org/optica/abstract.cfm?URI=optica-11-5-686},
   year = {2024},
   type = {Journal Article}
}

@article{RN469,
   author = {Li, G. Y. and Feng, X. and Yun, S. H.},
   title = {In Vivo Optical Coherence Elastography Unveils Spatial Variation of Human Corneal Stiffness},
   journal = {IEEE Transactions on Biomedical Engineering},
   volume = {71},
   number = {5},
   pages = {1418-1429},
   ISSN = {1558-2531},
   DOI = {10.1109/TBME.2023.3338086},
   year = {2024},
   type = {Journal Article}
}

@article{RN98,
   author = {Zvietcovich, Fernando and Pongchalee, Pornthep and Meemon, Panomsak and Rolland, Jannick P. and Parker, Kevin J.},
   title = {Reverberant 3D optical coherence elastography maps the elasticity of individual corneal layers},
   journal = {Nature Communications},
   volume = {10},
   number = {1},
   pages = {4895},
   ISSN = {2041-1723},
   DOI = {10.1038/s41467-019-12803-4},
   url = {https://doi.org/10.1038/s41467-019-12803-4},
   year = {2019},
   type = {Journal Article}
}

@article{RN875,
   author = {Ma, S. and Wang, R. and Qiu, S. and Li, R. and Yue, Q. and Sun, Q. and Chen, L. and Yan, F. and Yang, G. Z. and Feng, Y.},
   title = {MR Elastography With Optimization-Based Phase Unwrapping and Traveling Wave Expansion-Based Neural Network (TWENN)},
   journal = {IEEE Transactions on Medical Imaging},
   volume = {42},
   number = {9},
   pages = {2631-2642},
   ISSN = {1558-254X},
   DOI = {10.1109/TMI.2023.3261346},
   year = {2023},
   type = {Journal Article}
}

@article{RN1156,
   author = {Pillai, Eva K. and Franze, Kristian},
   title = {Mechanics in the nervous system: From development to disease},
   journal = {Neuron},
   volume = {112},
   number = {3},
   pages = {342-361},
   ISSN = {0896-6273},
   DOI = {https://doi.org/10.1016/j.neuron.2023.10.005},
   url = {https://www.sciencedirect.com/science/article/pii/S0896627323007596},
   year = {2024},
   type = {Journal Article}
}

@article{RN1397,
   author = {Xiao, Bailong},
   title = {Mechanisms of mechanotransduction and physiological roles of PIEZO channels},
   journal = {Nature Reviews Molecular Cell Biology},
   volume = {25},
   number = {11},
   pages = {886-903},
   ISSN = {1471-0080},
   DOI = {10.1038/s41580-024-00773-5},
   url = {https://doi.org/10.1038/s41580-024-00773-5},
   year = {2024},
   type = {Journal Article}
}

@article{RN1536,
   author = {Nia, Hadi T. and Munn, Lance L. and Jain, Rakesh K.},
   title = {Probing the physical hallmarks of cancer},
   journal = {Nature Methods},
   volume = {22},
   pages = {1800–1818},
   ISSN = {1548-7105},
   DOI = {10.1038/s41592-024-02564-4},
   url = {https://doi.org/10.1038/s41592-024-02564-4},
   year = {2025},
   type = {Journal Article}
}

@article{RN1741,
   author = {Boutouyrie, Pierre and Chowienczyk, Phil and Humphrey, Jay D. and Mitchell, Gary F.},
   title = {Arterial Stiffness and Cardiovascular Risk in Hypertension},
   journal = {Circulation Research},
   volume = {128},
   number = {7},
   pages = {864-886},
   DOI = {doi:10.1161/CIRCRESAHA.121.318061},
   url = {https://www.ahajournals.org/doi/abs/10.1161/CIRCRESAHA.121.318061},
   year = {2021},
   type = {Journal Article}
}

@article{RN1742,
   author = {Singh, Manmohan and Wu, Chen and Liu, Chih-Hao and Li, Jiasong and Schill, Alexander and Nair, Achuth and Larin, Kirill V.},
   title = {Phase-sensitive optical coherence elastography at 1.5 million A-Lines per second},
   journal = {Optics Letters},
   volume = {40},
   number = {11},
   pages = {2588-2591},
   DOI = {10.1364/OL.40.002588},
   url = {https://opg.optica.org/ol/abstract.cfm?URI=ol-40-11-2588},
   year = {2015},
   type = {Journal Article}
}

@article{RN1151,
   author = {Schmidt, Ginger and Bouma, Brett E. and Uribe-Patarroyo, Néstor},
   title = {Asynchronous, semi-reverberant elastography},
   journal = {Optica},
   volume = {11},
   number = {9},
   pages = {1285-1294},
   DOI = {10.1364/OPTICA.528507},
   url = {https://opg.optica.org/optica/abstract.cfm?URI=optica-11-9-1285},
   year = {2024},
   type = {Journal Article}
}

@article{RN872,
   author = {Yin, Z. and Li, G. Y. and Zhang, Z. and Zheng, Y. and Cao, Y.},
   title = {SWENet: A Physics-Informed Deep Neural Network (PINN) for Shear Wave Elastography},
   journal = {IEEE Transactions on Medical Imaging},
   volume = {43},
   number = {4},
   pages = {1434-1448},
   ISSN = {1558-254X},
   DOI = {10.1109/TMI.2023.3338178},
   year = {2024},
   type = {Journal Article}
}

@article{RN337,
   author = {Ramier, Antoine and Eltony, Amira M. and Chen, YiTong and Clouser, Fatima and Birkenfeld, Judith S. and Watts, Amy and Yun, Seok-Hyun},
   title = {In vivo measurement of shear modulus of the human cornea using optical coherence elastography},
   journal = {Scientific Reports},
   volume = {10},
   number = {1},
   pages = {17366},
   ISSN = {2045-2322},
   DOI = {10.1038/s41598-020-74383-4},
   url = {https://doi.org/10.1038/s41598-020-74383-4},
   year = {2020},
   type = {Journal Article}
}

@article{RN1755,
   author = {Ding, Weilian and Yang, Yang and Zhao, Yan and Jiang, Shichun and Cao, Yanping and Lu, Conghua},
   title = {Well-defined orthogonal surface wrinkles directed by the wrinkled boundary},
   journal = {Soft Matter},
   volume = {9},
   number = {14},
   pages = {3720-3726},
   ISSN = {1744-683X},
   DOI = {10.1039/C2SM27359D},
   url = {http://dx.doi.org/10.1039/C2SM27359D},
   year = {2013},
   type = {Journal Article}
}

@article{RN81,
   author = {Prevedel, Robert and Diz-Muñoz, Alba and Ruocco, Giancarlo and Antonacci, Giuseppe},
   title = {Brillouin microscopy: an emerging tool for mechanobiology},
   journal = {Nature Methods},
   volume = {16},
   number = {10},
   pages = {969-977},
   ISSN = {1548-7105},
   DOI = {10.1038/s41592-019-0543-3},
   url = {https://doi.org/10.1038/s41592-019-0543-3},
   year = {2019},
   type = {Journal Article}
}

@article{RN464,
  author  = {Ambrozi{\'n}ski, {\L}ukasz and Song, Shaozhen and
             Yoon, Soon Joon and Pelivanov, Ivan and Li, David and
             Gao, Liang and Shen, Tueng T. and Wang, Ruikang K. and
             O'Donnell, Matthew},
  title   = {Acoustic micro-tapping for non-contact {4D} imaging of tissue elasticity},
  journal = {Scientific Reports},
  volume  = {6},
  number  = {1},
  pages   = {38967},
  year    = {2016},
  doi     = {10.1038/srep38967},
  url     = {https://doi.org/10.1038/srep38967},
  issn    = {2045-2322}
}

@article{RN1756,
   author = {Burhan, Sazgar and Detrez, Nicolas and Rewerts, Katharina and Strenge, Paul and Buschschlüter, Steffen and Kren, Jessica and Hagel, Christian and Bonsanto, Matteo Mario and Brinkmann, Ralf and Huber, Robert},
   title = {Phase unwrapping for MHz optical coherence elastography and application to brain tumor tissue},
   journal = {Biomedical Optics Express},
   volume = {15},
   number = {2},
   pages = {1038-1058},
   DOI = {10.1364/BOE.510020},
   url = {https://opg.optica.org/boe/abstract.cfm?URI=boe-15-2-1038},
   year = {2024},
   type = {Journal Article}
}

@article{RN1231,
   author = {Kirby, Mitchell and Zhou, Kanheng and Pitre, John and Gao, Liang and Li, David and Pelivanov, Ivan and Song, Shaozhen and Li, Chunhui and Huang, Zhihong and Shen, Tueng and Wang, Ruikang and O'Donnell, Matthew},
   title = {Spatial resolution in dynamic optical coherence elastography},
   journal = {Journal of Biomedical Optics},
   volume = {24},
   number = {9},
   pages = {096006},
   url = {https://doi.org/10.1117/1.JBO.24.9.096006},
   year = {2019},
   type = {Journal Article}
}

\end{document}